\documentclass[a4paper]{jpconf}
\usepackage{graphicx}
\usepackage{amsmath}
\usepackage{amsfonts}

\newcommand{\beq}{\begin{eqnarray}}
\newcommand{\eeq}{\end{eqnarray}}

\begin{document}

\title{Transversity studies with protons and light nuclei}

\author{Sergio Scopetta}

\address{
Dipartimento di Fisica, Universit\`a degli Studi
di Perugia, \\
and INFN, sezione di Perugia, \\
via A. Pascoli, 06100 Perugia, Italy
}

\ead{sergio.scopetta@pg.infn.it}

\begin{abstract}
A general formalism to evaluate time-reversal odd 
transverse momentum dependent parton distributions
(T-odd TMDS) 
is reviewed and applied to the calculation of the leading-twist 
quantities,
i.e., the Sivers and the Boer-Mulders functions. 
Two different models of the proton structure, 
namely 
a non relativistic constituent quark model and the MIT bag model,
have been used. The results obtained in both frameworks fulfill
the Burkardt sum rule to a large extent. 
The calculation of
nuclear effects in the extraction of neutron
single spin asymmetries in semi-inclusive deep inelastic scattering
off transversely polarized $^3$He is also illustrated. 
In the kinematics of JLab, 
it is found that the nuclear effects described by an 
Impulse Approximation approach are under theoretical control.
\end{abstract}

\section{Introduction}

The partonic structure of transversely polarized hadrons
is still rather poorly known \cite{bdr}.
As a matter of facts, despite its leading-twist character,
the transversity distribution is chiral-odd in 
nature and not
accessible therefore in inclusive esperiments.
Due to going-on and forthcoming important measurements
in various Laboratories, this issue is nevertheless one of the most
widely studied by the hadronic Physics Community,
in particular by the Italian one, which
is providing crucial contributions to a better
understanding of this fascinating subject (see, for example,
Refs. \cite{ferrara} and references therein).

Semi-inclusive deep inelastic scattering (SIDIS)
is one of the proposed
processes to access the parton structure
of transversely polarized hadrons. 
The theoretical description of semi-inclusive processes 
implies a more complicated formalism,
accounting for the transverse motion of the quarks
in the target \cite{sidis,D'Alesio:2007jt,Pasquini:2008ax}.
In particular, the non-perturbative effects of
the intrinsic transverse momentum $\vec{k}_{T}$ of the quarks inside 
the nucleon may induce significant hadron azimuthal asymmetries 
\cite{Mulders:1995dh, Cahn:1978se}.

The Sivers and the Boer-Mulders functions were defined in this scenario
\cite{sivers, Boer:1997nt}. 
Transverse Momentum Dependent pdfs (TMDs) are the set of functions 
that depend on the intrinsic 
transverse momentum of the quark, in addition to the dependences,
typical of the PDs, on
the Bjorken variable and on the momentum transfer $Q^2$.
Their number is fixed counting the scalar quantities
allowed by hermiticity, parity and time-reversal invariance. 
However, the existence of final state interactions (FSI) allows for 
time-reversal odd functions~\cite{Brodsky:2002cx}. In effect, 
by relaxing this constraint, one defines two additional functions, 
namely, the Sivers and the Boer-Mulders (BM) functions.
In SIDIS, the Sivers function is involved in the description of the
single spin asymmetry, measured when an unpolarized beam
scatters transeversely polarized targets.
The single spin asymmetries are obtained constructing the difference of
semi-inclusive cross-sections with different transverse polarization
of the target with respect to the momentum transfer.
The BM function appears in the
azimuthal asymmetry in unpolarized SIDIS. The latter
object refers to the detection
of the produced hadron at different angles with respect
to the plane containing the momentum transfer and the
hadron momentum.
All the involved quantities have 
been conventionally defined in Ref. \cite{trento}.
According to the latter convention,
the Sivers function, $f_{1T}^{\perp {q} } (x, k_T)$ \cite{sivers}, 
and the Boer-Mulders function,  $h_{1}^{\perp {q} } (x, k_T)$ 
\cite{Boer:1997nt}, are formally defined through the following 
expressions\footnote{$a^\pm = (a_0 \pm a_3)/ \sqrt{2}$.}:
\begin{eqnarray}
f_{1T}^{\perp {q}} (x, {k_T} ) 
& = &- {M \over 2k_x} \,\int \frac{d\xi^- d^2\vec{\xi}_T}{(2\pi)^3} \ e^{-i(xp^+ \xi^- -\vec{k}_T\cdot \vec{\xi}_T)} \,\nonumber\\
& \times &
{1 \over 2} \sum_{S_y=-1,1} \, S_y \, \langle P S_y\vert \overline \psi_{q} (\xi^-, \vec{\xi}_T)\, {\cal L}^{\dagger}_{\vec{\xi}_T}(\infty, \xi^-)\, 
\gamma^+\,{\cal L}_0(\infty,0) \psi_{q}(0,0) \vert P S_y\rangle \, 
\nonumber \\
& + & \mbox{h.c.} \, ,
\label{siv-def-op}
\eeq
taking the proton polarized  along the $y$ axis; and
\beq
h_{1}^{\perp {q}} (x, k_T) 
& = & - {M \over 2 k_x} \,\int \frac{d\xi^- d^2\vec{\xi}_T}{(2\pi)^3} \ e^{-i(xp^+ \xi^- -\vec{k}_T\cdot \vec{\xi}_T)} \,\nonumber\\
& \times &
{1 \over 2} \sum_{S_z=-1,1} \langle P S_z\vert \overline \psi_{q} (\xi^-, \vec{\xi}_T)\,  {\cal L}^{\dagger}_{\vec{\xi}_T}(\infty, \xi^-)\, 
\gamma^+\gamma^2\gamma_5\,{\cal L}_0(\infty,0) \psi_{q}(0,0) \vert P S_z\rangle \, 
\nonumber \\
& + & \mbox{h.c.} \, ,
\label{bm-def-op}
\eeq
where $ {\vec S} $ is the spin of the target hadron. The normalization of the covariant spin vector is $S^2 = -1$,
$M$ is the target mass, $\psi_{q}(\xi)$ is the quark field
and  ${\cal L}_{\vec{\xi}_T}$ is the gauge link.\footnote{The gauge link is defined as
$
 {\cal L}_{\vec{\xi}_T}(\infty, \xi^-)= {\cal P} \mbox{exp}\left( -ig\, \int_{\xi^-}^{\infty} \, A^+(\eta^-,\vec{\xi}_T)\, d\eta^-\right)\,, 
$
where $g$ is the strong coupling constant. This definition holds in 
covariant (non singular) gauges, and in SIDIS processes, as the definition 
of T-odd  TMDs is process dependent.}
The gauge link contains a scaling contribution which makes the T-odd 
TMDs non vanishing in the Bjorken limit, as it has been shown
in Refs. \cite{brodhoy,coll2,jiyu}.

The difference between the two functions is clearly  
seen comparing  Eq. (\ref{siv-def-op}) and Eq. (\ref{bm-def-op}). 
The BM function counts transversely  polarized quarks, 
hence the Dirac operator $\gamma^+\gamma^2\gamma_5$ in Eq. (\ref{bm-def-op}), 
in an unpolarized proton. On the other hand, the Sivers function counts 
the unpolarized quarks, hence the Dirac  operator $\gamma^+$ in Eq. 
(\ref{siv-def-op}), in a transversly polarized proton, denoted by
the transverse component $S_y$ in the proton state in Eq. (\ref{siv-def-op}). 
If there were no scaling contribution of the gauge link, 
the two T-odd functions, $f_{1T}^{\perp {q} } (x, k_T)$ 
and $h_{1}^{\perp {q} } (x, k_T)$, would be identically zero.

\section{Quark model calculations}

Several quark model calculations
of the Sivers and BM functions have been
performed in the past years
(see, i.e., Refs. \cite{yuan,Gamberg:2007wm,Bacchetta:2008af,
Pasquini:2010af})
and their phenomenology keeps attracting interest,
as it is demonstrated by very recent important
contributions \cite{baro,boernew,bacnew}.
\begin{figure}
\begin{center}
\includegraphics[height=3.5cm]{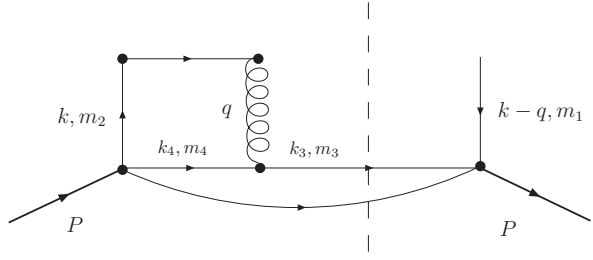}%
\caption{
The process taken into account, together with its
h.c., in the present evaluation of the T-odd TMDs.
}
\label{feyn}%
\end{center}
\end{figure}

The general formalism for the evaluation of time-reversal odd TMDs
in quark models, presented in Ref.~\cite{Courtoy:2008vi, Courtoy:2008dn, Courtoy:2009pc}, is reviewed here below. 
It is based on an 
impulse approximation analysis, where the Final State Interactions 
are introduced through a One Gluon Exchange mechanism, 
as depicted in Fig.~\ref{feyn}.
The used quark models, i.e., the
simplest version of the MIT bag model ~\cite{Jaffe:1974nj},
and a non relativistic constituent quark model (NRCQM),
do not contain explicit gluonic degrees of freedom.
In the Sivers and BM functions 
the Dirac operators determine the spin structure 
of the interaction, $V_{m_1m_2m_3m_4}$ between the quarks
with initial spins $m_2$ and $m_4$ and final spins $m_1$ and $m_3$, 
respectively.
For instance, in the MIT bag model~\cite{Jaffe:1974nj}, one gets,
for the Sivers [BM] function
\beq
f_{1T} [h_{1}]^{\perp {q}} (x, k_T) &=& -2\Im[\,]\,  
i g^2\,\frac{M\, E_P}{k_{x}}\, 
\int \frac{d^2 \vec{q}_T}{(2\pi)^5}
\int \frac{d^3 k_3}{(2\pi)^3}
\nonumber \\
& \times &
\sum_{m_1,m_2,m_3, m_4} 
C_{{q} f[h]}^{m_1m_2, m_3m_4}\, 
V(\vec{k}, \vec{k}_3, \vec{q}_T)_{m_1m_2, m_3m_4}
 \quad.
\label{bm-bag-ready}
\eeq
The imaginary part is taken only for the Sivers function calculation, 
performed changing the transversely polarized
nucleon states from a transversity to a
helicity basis. 
The coupling of the spins at the quark level appears clearly 
from the expressions
  \beq
  C_{{q} f[h]}^{m_1m_2, m_3m_4} &=&{1 \over 2} \sum_{S}\,C_{{q}S}^{m_1m_2, m_3m_4}\nonumber\\
  &=&\sum_{\beta} \,T^a_{ij}T^a_{kl}\,\langle P S_z=1|  b_{{q} m_1}^{i\dagger}b_{{q} m_2}^{j}b_{\beta m_3}^{k\dagger}b_{\beta m_4}^{l} |P  S_z=-1\rangle\nonumber\\
&&\left[{1 \over 2} \sum_{S_z=1,-1}\,\sum_{\beta} \,T^a_{ij}T^a_{kl}\,\langle P S_z|  b_{{q} m_1}^{i\dagger}b_{{q} m_2}^{j}b_{\beta m_3}^{k\dagger}b_{\beta m_4}^{l} |P  S_z\rangle\right]
\quad,
\label{me}  
\eeq
whose calculation has been performed assuming $SU(6)$ symmetry for the proton 
state,  
for ${q}=u,d$. The interaction $V(\vec{k},\vec{k}_3, 
\vec{q}_T)$ has been evaluated using
the properly normalized fields for the quark in the bag \cite{Jaffe:1974nj},  
given in terms of the quark wave function in momentum
space, which reads
\begin{eqnarray}
\varphi_m(\vec{k})&=&i\, \sqrt{4\pi}\, N\, R_0^3 
\begin{pmatrix}
 t_0(|\vec{k}|) \chi_m
\\
{\vec{\sigma} \cdot \hat{k}}\,t_1(|\vec{k}|)\, \chi_m
\end{pmatrix} \quad,
\label{bagwf}
\end{eqnarray}
with the normalization factor $N$ given in Ref.
\cite{Jaffe:1974nj}
The interaction is then
\beq
V(\vec{k}, \vec{k}_3, \vec{q}_T)_{m_1m_2, m_3m_4} &=&
\frac{1}{q^2}\, \varphi^{\dagger}_{m_1} (\vec{k}-\vec{q}_T)\,\gamma^0 \gamma^+
\Gamma^{f[h]}\,
 \varphi_{m_2} (\vec{k})
\varphi^{\dagger}_{m_3} ( \vec{k}_{3})\,\gamma^0 \gamma^+\, \varphi_{m_4} 
(\vec{k}_{3}-\vec{q}_T)\,,
\label{int-bag}
\eeq
where $\Gamma^{f[h]}=1$ or $\gamma^2\gamma_5$ for, respectively, 
the  Sivers and the Boer-Mulders  function.
The expressions in the NRCQM are similar to the one in the bag model. 
They read, being $\Psi$ the intrinsic proton wave function,
\beq
f_{1T} [h_{1}]^{\perp {q}} (x, k_T)&=&-2 i g^2\,\frac{M\, E_P}{k_{x}}\, 
\int \frac{d^2 \vec{q}_T}{(2\pi)^5}
\,
\sum_{m_1,m_2,m_3, m_4}\nonumber\\
&&\int \,d\vec{k}_1 \, d\vec{k}_3\,(2\pi)^3 \delta(k_1^+-xP^+) \delta(\vec{k}_{1\perp}+\vec{q}_{\perp}-\vec{k}_{\perp})\nonumber\\
&&{1 \over 2} \sum_{S}\,\sum_{\beta}\sum_{ijkl}
\delta_{\mbox{\tiny QN}} \,\,
\Psi^{\dagger}_{r \, S}
\left ( \vec k_1, \{m_1,i,{q}\}; \, \vec k_3, \{m_3,k,\beta \};
\, - \vec k_3 - \vec k_1,  m_n  \right )\,\,
\nonumber \\
& \times &
T^a_{ij}  T^a_{kl}
V(\vec k_1, \vec k_3, \vec q)_{m_1m_2, m_3m_4}
\nonumber \\
& \times &
\Psi_{r \, S}
\left (\vec k_1 + \vec q, \{m_2,j,{q}\}; \, \vec k_3 -
\vec q, \{m_4,l,\beta \};
\, - \vec k_3 - \vec k_1,  m_n  \right )
 \, ,
 \nonumber\\
\label{bm-cqm-ready}
\eeq
where $\sum_{S}=\sum_{S_z=-1,1}$ or $\sum_{S_y=-1,1}\,S_y$ for the BM and 
the Sivers function, respectively.
The interaction here reads
\beq
V(\vec k_1, \vec k_3, \vec q)_{m_1m_2m_3 m_4}
&=& \frac{1}{q^2}\, \bar u_{m_1}(\vec{k}_1)
\,\gamma^+\Gamma^{f[h]}\,
u_{m_2}(\vec{k}_1 + \vec q )
\,  \bar u_{m_3}(\vec{k}_3)
\,\gamma^+\,
u_{m_4}(\vec{k}_3-\vec{q})
\quad,
\label{int}
\eeq
with $u(\vec{k})$ the four-spinor of the free quark states, arising
from the Impulse Approximation analysis

In Refs.~\cite{Courtoy:2008vi, Courtoy:2008dn, Courtoy:2009pc}, 
the T-odd TMDs 
have been evaluated in both the MIT bag model and the NRCQM
assuming initially
an $SU(6)$ symmetry for the proton. 

In the former case, the qualitative results are  
understood calculating the coefficients (\ref{me}),
in terms of which it is possible to realize what happens 
to the quark spins in the FSI process 
in a perfectly transparent way. 
In the latter case,  one first has to re-express Eq.~(\ref{bm-cqm-ready}) 
in terms of the proton state. In spectroscopic notation and with the 
Jacobi coordinates, one has, using SU(6) symmetry
\beq
 | ^2 S_{1/2} \rangle_S&=& \frac{
 e^{-\left(k^2_{\rho} +k_{\lambda}^2\right)/\alpha^2}
 }{\pi^{3/2} \alpha^3} | \chi \rangle_S~,
\label{wf}
\eeq
where $  | \chi \rangle_S $ 
is the standard $SU(6)$ vector describing the spin-flavor structure 
of the proton.
In the bag model calculation, a non-zero Sivers 
function arises through the interference between
the upper and lower components of the wave function.
Something similar happens in the CQM. This time the interference
is between the upper and lower components of the free
spinors appearing as a consequence of the use of the Impulse Approximation
(see 
Eqs.~(\ref{int-bag}, \ref{int}) for the interaction and
Refs. 
\cite{Courtoy:2008vi, Courtoy:2008dn, Courtoy:2009pc}
for details).
Two different spin combinations contribute to the Sivers function.
One of them comes from the spin-flipping of the 
quark interacting with the photon, i.e. the  $Y$ term in the MIT bag 
calculation
\begin{eqnarray}
f_{1T}^{\perp {q}} (x, k_T)&=&\frac{g^2}{2}\frac{ME_P}{k^x} \,C^2
\int
\frac{d^2q_\perp}{(2\pi)^2}\frac{1}{q^2} [ C_{q}^{-+} Y(\vec
q_\perp,k_T) + C_{q}^{+-} U(\vec q_\perp,k_T)]~,\label{sivf}
\end{eqnarray}
with $C$ a normalization factor, $C_{q}$ a weighting spin-flavor-color 
factor resulting from the matrix elements~(\ref{me}) and where $Y/U(\vec
q_\perp,k_T)$ include the momentum dependent part.\footnote{See Eqs.~(8-9) 
of Ref.~\cite{Courtoy:2008dn}.}

On the other hand, there are more spin combinations for 
the BM function.\footnote{See Eq.~(13) of Ref.~\cite{Courtoy:2009pc}.
} The first reason is that both non-flipping and double-flipping 
terms are important. The second reason is the sum over the two spin 
states, i.e. $S_z=-1,1$.
Due to the spin-flavor-color coefficients, i.e., due to the
$SU(6)$ symmetry assumption, the non-flipping term is more important
than the double-flipping contribution.
In effect, the latters are governed by the product of the
two lower components of the bag wave function  which encodes 
the most relativistic contribution
arising in the MIT bag model.
They turn out to be a few orders of magnitude smaller
than the dominant ones, arising from the interference between
the upper and lower parts of the bag wave function.
This  also happens if a proper non relativistic reduction of the gauge link, 
suitable for CQM calculations, is performed, justifying then the non 
relativistic approximation. 

In Fig. 2 the ``first moments'' of the Sivers and the 
Boer-Mulders functions,
i.e. the quantities
\begin{equation}
f_{1T}^{\perp (1) {\cal Q} } (x)
= \int {d^2 \vec k_T}  { k_T^2 \over 2 M^2}
f_{1T}^{\perp {\cal Q}} (x, {k_T} )~,
\label{momf}
\end{equation}
are shown for $u$ and $d$ quarks in both the CQM and 
MIT bag model, using $\alpha_s(\mu_0^2)/(4 \pi) \simeq 0.13$
~\cite{Traini:1997jz}. 

\begin{figure}[t]
\begin{minipage}{18pc}
\hspace{-6pc}
\includegraphics[width=30pc]{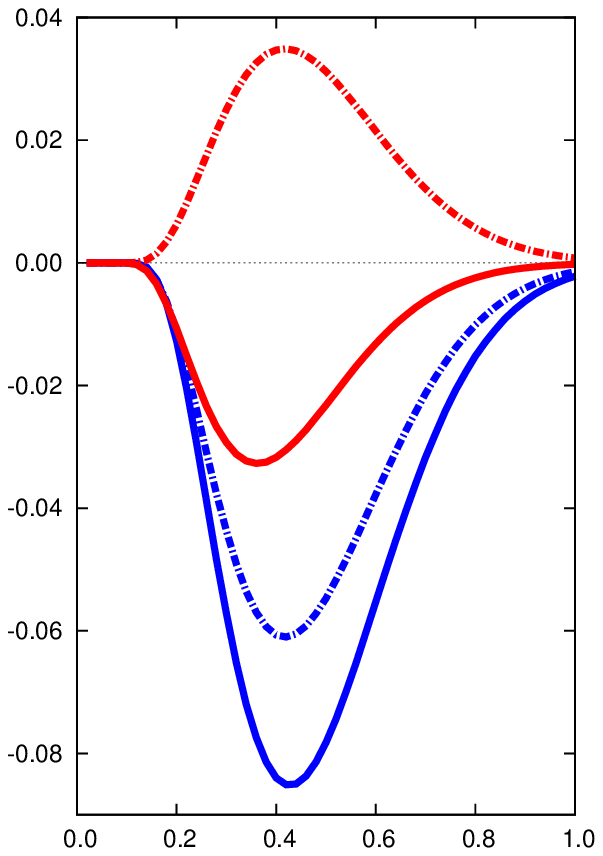}
\begin{center}
\hspace{2pc} x
\end{center}
\end{minipage}\hspace{-6pc}%
\begin{minipage}{18pc}
\includegraphics[width=30pc]{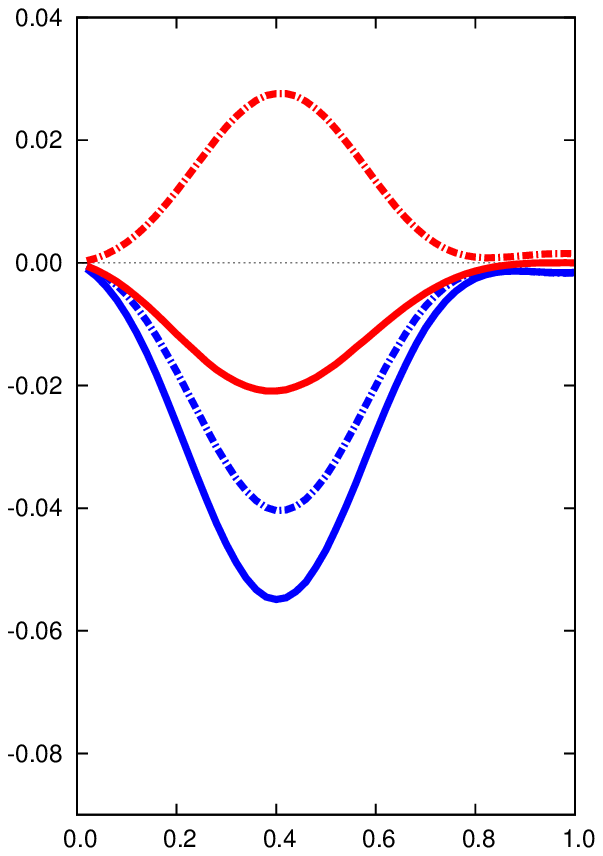}
\begin{center}
\hspace{14pc} x
\end{center}
\end{minipage} 
\caption{
Comparison of the 
moments, $f(h)_{1T}^{\perp (1) u(d) }(x) $, Eq. (\ref{momf}),
of the
Sivers (red) and Boer-Mulders (blue) functions in the 
SU(6) NR Constituent Quark Model (left) and the MIT bag model (right). 
Short-dashed curves represent the $d$-quark distributions; 
full curves the $u$-quark ones. 
}
\end{figure}
\begin{figure}
\begin{center}
\includegraphics[width=.75\textwidth]{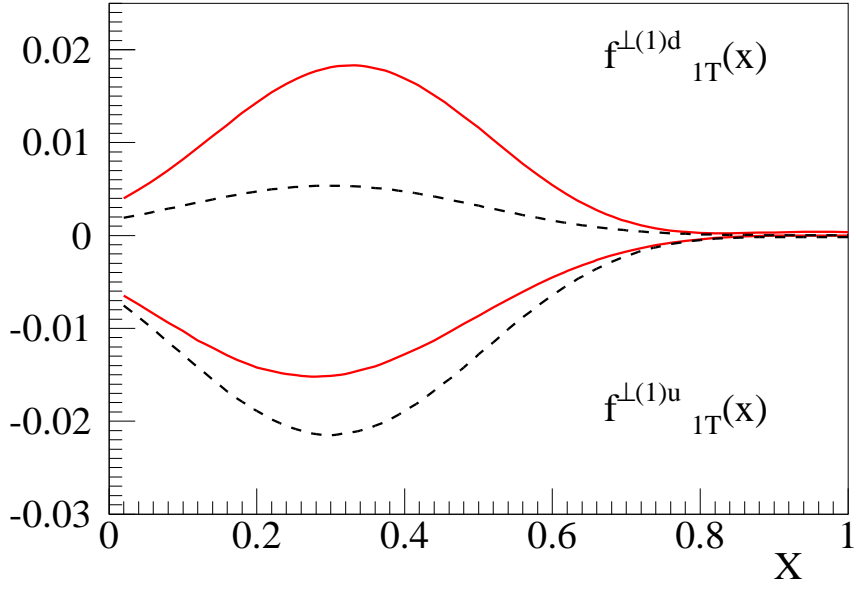}%
\vskip -7. cm
\caption{
The quantity $f_{1T}^{\perp (1) u(d) }(x) $, Eq. (\ref{momf}).
Dashed curve: the results in the MIT bag model at  $\mu_0^2$,
considering that a spin-flip can occurr only for the
interacting quark. 
Full curve: the same, evaluated properly taking into account
also the possible spin-flip of the spectator quark.
}
\end{center}
\end{figure}
In Fig. 3 it is shown, in the MIT bag model case, the effect
of neglecting the possible contribution of a spin-flip of the
spectator quark (the one with initial
spin projection $m_4$ in Fig. 1), as was done in a previous calculation
\cite{yuan}. The proper inclusion of this term is crucial
for the fulfillment of the Burkart Sum Rule, as it is explained here below. 

The above-described formalism has been easily extended, in the NR case, 
to the model of Isgur and Karl \cite{ik}, allowing for a weak
SU(6) breaking. The detailed procedure and
the final expressions of the Sivers function in this model can be found
in Ref. \cite{Courtoy:2008vi}.
\begin{figure}
\includegraphics[width=.49\textwidth]{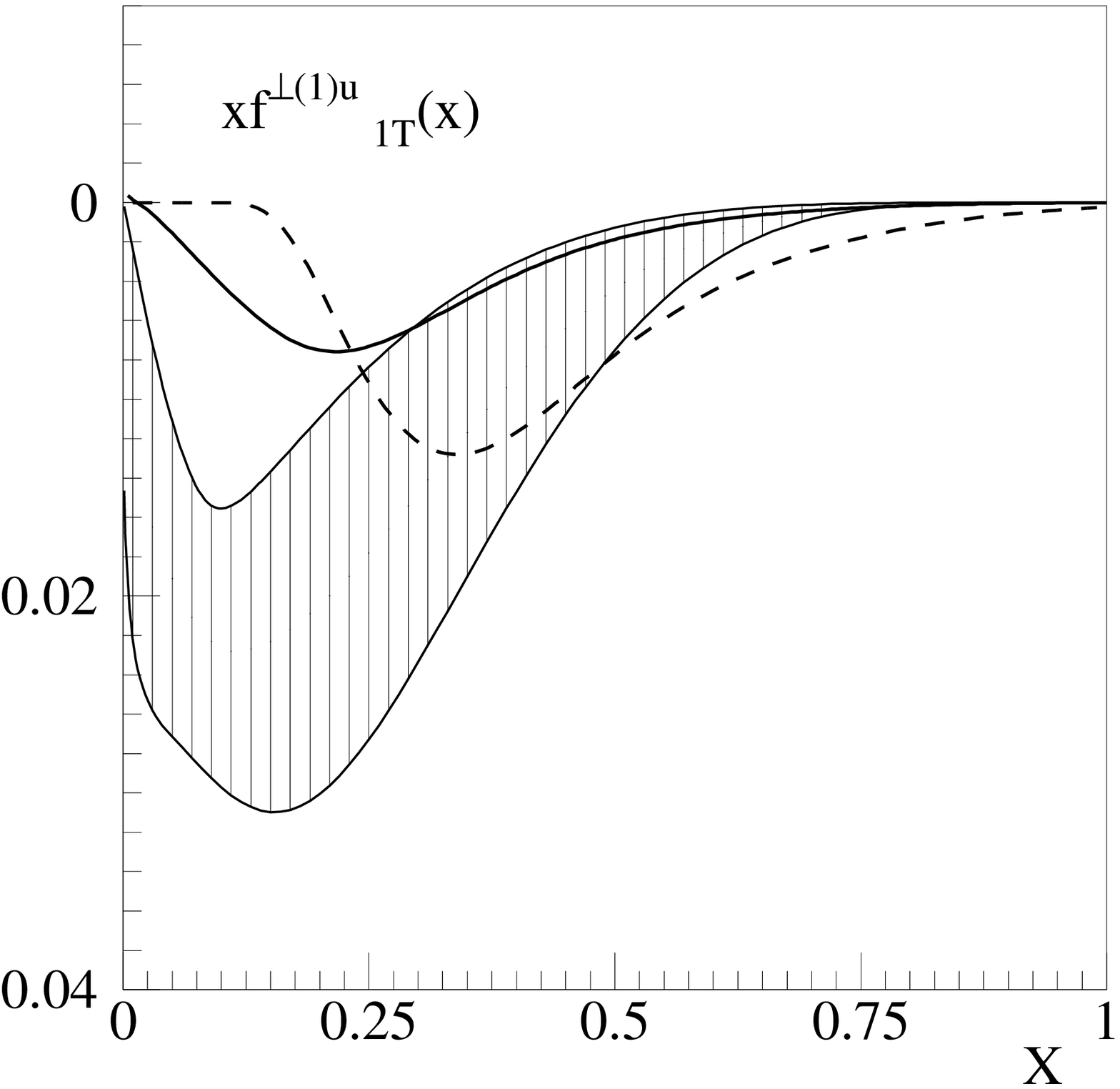}
\includegraphics[width=.49\textwidth]{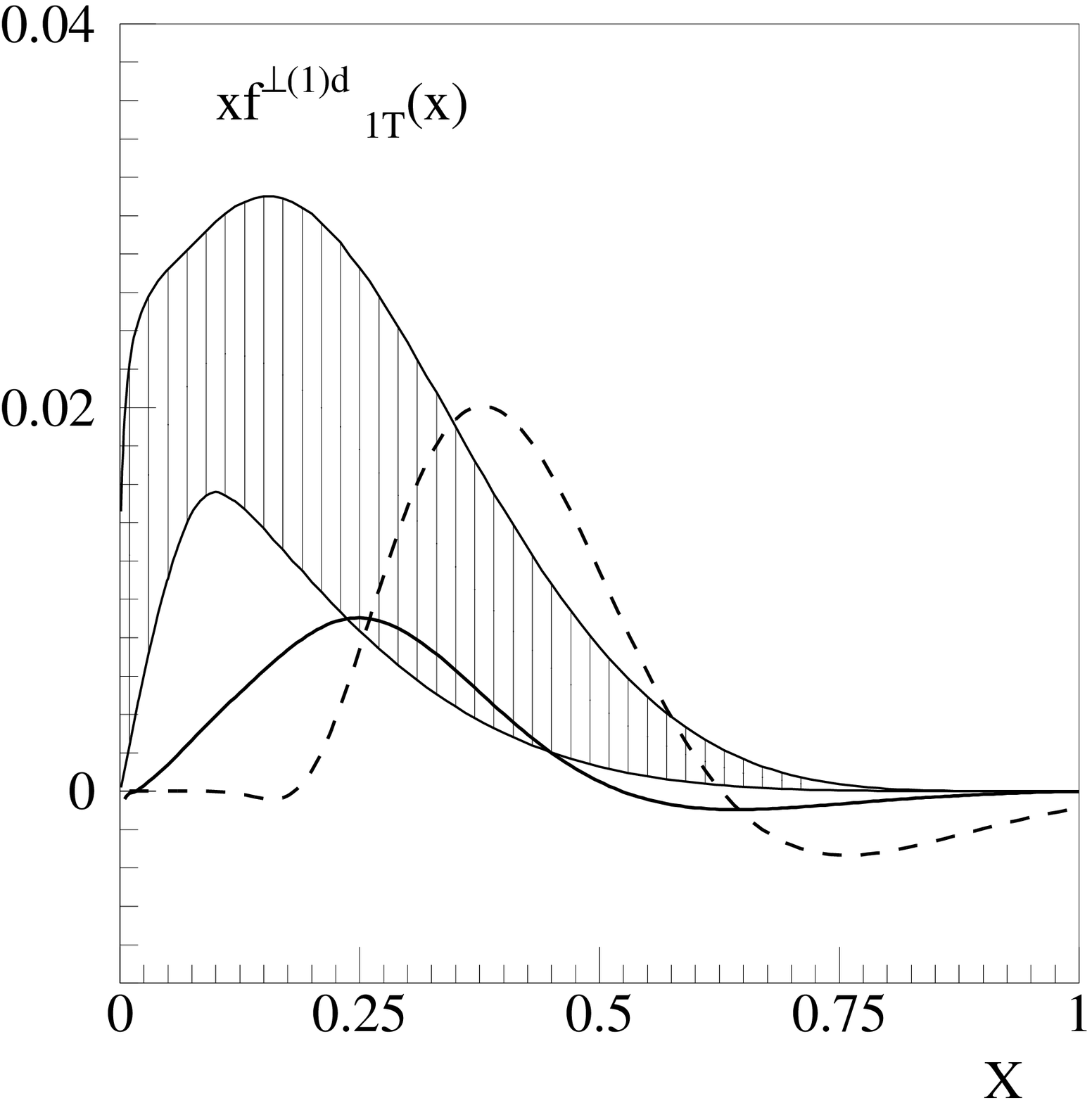}
\caption{
Left (right): the quantity $f_{1T}^{\perp (1) u(d) }(x) $, Eq. (\ref{momf}).
Dashed curve: the result in the Isgur and Karl model at  $\mu_0^2$.
Full curve:  the evolved distribution at NLO.
Patterned area: the $1 - \sigma$ range of the best fit 
of the HERMES data proposed in Ref. \cite{coll3}.
}
\end{figure}
To evaluate numerically the Sivers and BM functions, 
the strong coupling constant $g$ (i.e.
$\alpha_s(Q^2)$) has to be fixed.
The prescription of Ref. 
~\cite{Traini:1997jz}. 
is used to fix $\mu_0^2$, according to the
amount of momentum carried by the valence quarks in the model.
Here, assuming that all the gluons and sea pairs in the proton
are produced perturbatively according to NLO evolution equations,
in order to have $\simeq 55 \% $ of the momentum
carried by the valence quarks at a scale of 0.34 GeV$^2$ 
one finds
that $\mu_0^2 \simeq 0.1$ GeV$^2$
if $\Lambda_{QCD}^{NLO} \simeq 0.24$ GeV.
This yields $\alpha_s(\mu_0^2)/(4 \pi) \simeq 0.13$ 
~\cite{Traini:1997jz}. 
The results of the present approach
for the first  moments of the Sivers function
are given
by the dashed curves in Fig. 4, where
they are compared
with a parameterization of the HERMES data,
taken at $Q^2=2.5$ GeV$^2$.
The patterned area represents the $1-\sigma$ range
of the best fit proposed in Ref. \cite{coll3}.
The results are close, in magnitude, to the data,
although they have a different shape: the maximum (minimum)
is predicted at larger values of $x$.
Actually $\mu_0^2$ is much lower, $Q^2 =2.5$ GeV$^2$. 
For a proper comparison, the QCD
evolution from the model scale to the experimental one
would be necessary. 
Unfortunately, the Sivers function is a TMD PDs and the evolution
of this class of functions is, to a large
extent, unknown.
In order to have an indication of the effect of the
evolution, 
a NLO evolution of the model
results has been performed, assuming, for $f_{1T}^{\perp (1) {\cal Q} } (x)$,
the same anomalous dimensions of the unpolarized PDFs.
From the final result (full curve in Fig. 2),  one can see
that the  agreement with data
improves dramatically and the
trend is reasonably reproduced at least for $x \ge 0.2$.
Although the performed evolution is not exact, 
the procedure highlights the necessity 
of evolving the model results 
to the experiment scale and it suggests 
that the present results could be
consistent with data,  
still affected by large errors.
A discussion concerning 
the use of perturbative methods to relate the scale of
model calculations to the scale of experimental data,
and its impact on the calculation of T-odd TMDs,
has been recently presented in Ref. 
\cite{Courtoy:2011mf}
The deep understanding of the QCD evolution of TMDs is progressing fast
\cite{Ceccopieri:2007ek,Kang:2008ey,Zhou:2008mz,
Vogelsang:2009pj,Cherednikov:2010uy}.

Let us see now how the results of the calculation
compare with the Burkardt sum rule
\cite{Burkardt:2004ur}, 
which follows from general principles. 
The Burkardt Sum Rule (BSR) states that, 
for a proton  polarized in the positive $y$ direction,
$\sum_{{\cal Q}=u,d} \langle k_x^{\cal{Q}} \rangle = 0$
with
\begin{equation}
\langle k_x^{\cal{Q}} \rangle = - \int_0^1 d x \int d \vec k_T
{k_x^2 \over M}  f_{1T}^{\perp \cal{Q}} (x, {k_T} )~,
\label{burs}
\end{equation}
and must be satisfied at any scale.
Within our scheme, at the scale of the model, it is found
$\langle k_x^{u} \rangle = 10.85$ MeV,
$\langle k_x^{d} \rangle = - 11.25$ MeV and, 
in order to have an estimate
of the quality of the agreement of our results with
the sum rule, we define the ratio
$r= 
| \langle k_x^{d} \rangle+
\langle k_x^{u}\rangle | /
| \langle k_x^{d} \rangle-
\langle k_x^{u} \rangle | $
obtaining $r \simeq 0.02$, so that one can say that the calculation
fulfills the BSR to a precision of a few percent.
One should notice that the agreement which is found
is better than that found in previous model calculations,
where the Sivers function
for the flavor $u$ was found to be proportional to that for the flavor $d$.
For the MIT bag model, where the translational invariance
is broken and the wave functions are not exact momentum
eigenstates, the situatuion is slightly worse
and the Sum Rule turns out to be violated of $\simeq$ 5 \%
(one should realize anyway that, in a previous calculation
without spin flip of the spectator quark, the violation
was found to be $\simeq$ 60 \%).

The recent calculation of Ref. \cite{Pasquini:2010af}, using
overlaps of Light Cone Wave Functions in the Light Front of Dynamics,
where the number of particles and the momentum sum rules
are satisfied at the same time, fulfills the Burkart Sume Rule exactly,
showing the importance of using exact momentum eigenstates
in checking the transverse momentum of the quarks.

\section{The Sivers function from neutron
($^3$He) targets}

The experimental scenario which arises from the analysis
of SIDIS off transversely polarized
proton and deuteron targets is puzzling \cite{hermes,compass}.
\begin{figure}
\includegraphics[width=.49\textwidth]{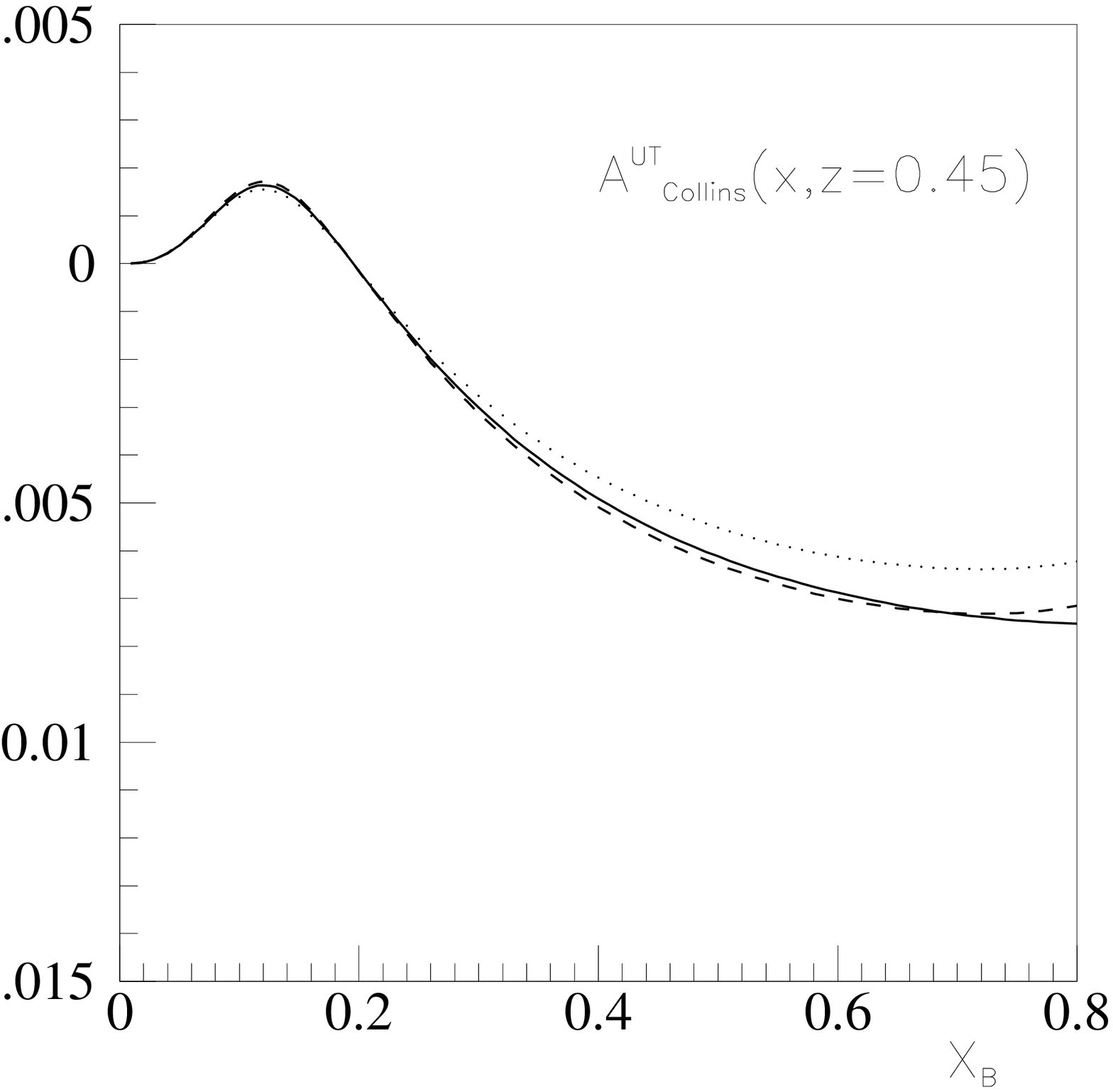}
\includegraphics[width=.49\textwidth]{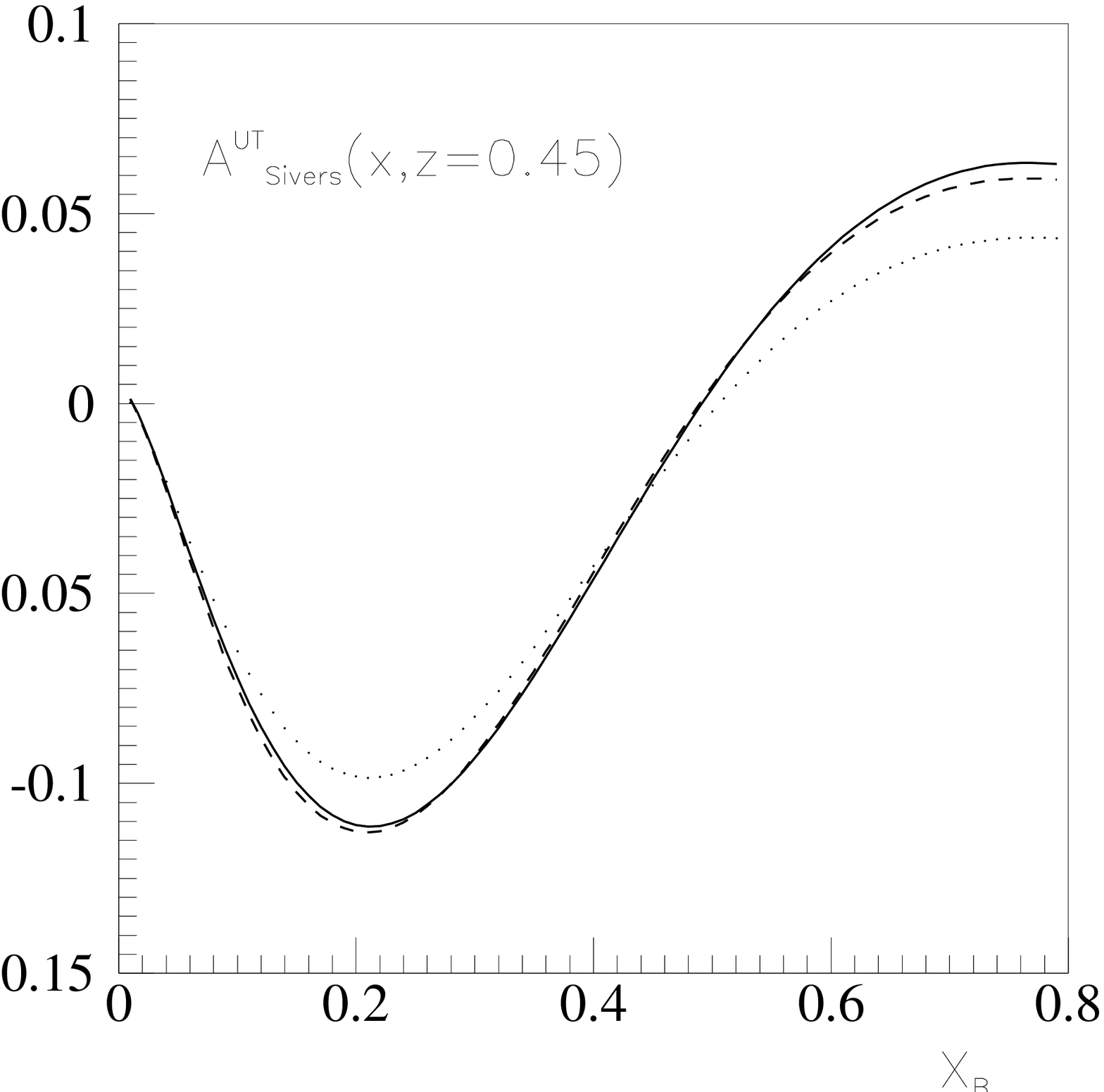}
\caption{Left (right):
the model neutron Collins (Sivers) asymmetry for
$\pi^-$ production
(full) in JLab kinematics, and the one extracted
from the full calculation taking into account
the $p_p$
(dashed), or neglecting it (dotted). 
The results are shown for {$z$}=0.45 and
$Q^2= 2.2$ GeV$^2$, typical values
in the kinematics of the JLab experiments.}
\end{figure}
\begin{figure}
\includegraphics[width=.49\textwidth,height=14.5cm]{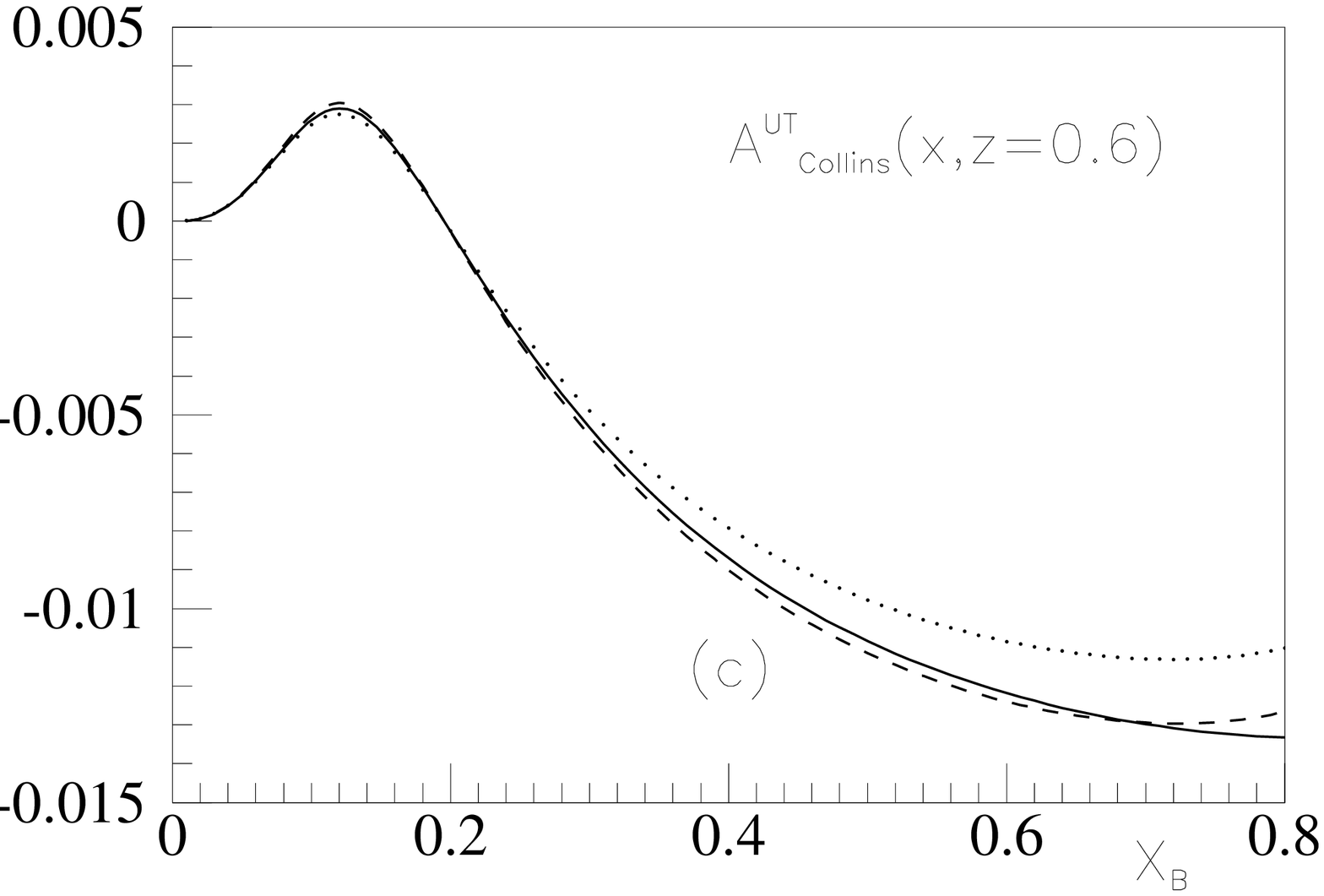}
\includegraphics[width=.49\textwidth,height=14.5cm]{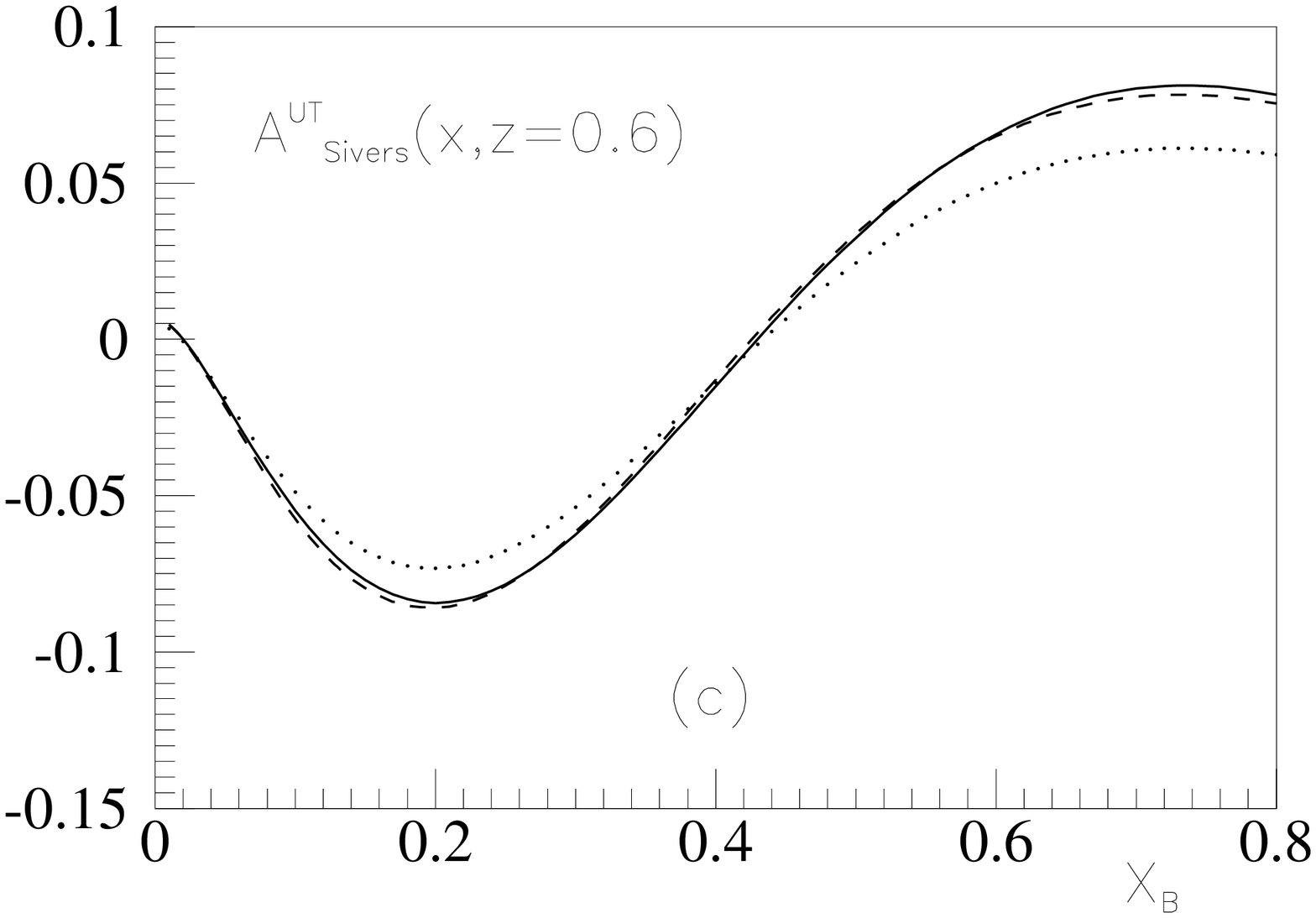}
\vskip -7. cm
\caption{Left (right)
The same as in Fig. 5, but at 
{$z$}=0.6, the maximum value accessed 
in the JLab experiments.}
\end{figure}
With the aim at extracting the neutron information
to shed some light on the problem,
a measurement of SIDIS
off transversely polarized $^3$He has been addressed \cite{bro},
and an experiment, planned to measure azimuthal asymmetries
in the production of leading $\pi^\pm$  from transversely
polarized $^3$He, has been just completed at Jefferson Lab, 
with a beam energy of 6 GeV
\cite{prljlab}. Another experiment
will be performed after the 12 GeV upgrade of JLab \cite{jlab12,alessio}.
Here, a realistic analysis of SIDIS
off transversely polarized $^3$He, presented in Ref.
\cite{mio}, is sommarized.
The formal expressions
of the
Collins and Sivers contributions to the azimuthal
Single Spin Asymmetry (SSA) for the production
of leading pions off $^3$He have been derived, in impulse approximation (IA),
including also the initial transverse momentum of the struck quark.
The final equations are rather involved and they are not
reported here. They can be found in \cite{mio}.
The same quantities have been then evaluated
in the kinematics of the planned 
JLab experiments.
Wave functions \cite{pisa} obtained within
the AV18 interaction \cite{av18} have been used for a realistic
description of the nuclear dynamics,
using overlap integrals evaluated in Ref. \cite{over},
and the nucleon structure has been described
by proper parameterizations of data \cite{ans}
or suitable model calculations
\cite{model}.
The crucial issue of extracting
the neutron information from $^3$He data
will be now thoroughly discussed. 
As a matter of facts,
a model independent procedure, based
on the realistic evaluation
of the proton and neutron polarizations in $^3$He
\cite{old}, called respectively $p_p$ and $p_n$ in the following,
is widely used in inclusive 
DIS to take into account effectively  
the momentum and energy distributions
of the polarized bound nucleons in $^3$He.
It is found that the same extraction technique
can be applied also in the 
kinematics of the proposed experiments, although
fragmentation functions, not only parton
distributions, are involved, as it can be seen
in Figs. 5 and 6. In these figures,
the free neutron asymmetry used as a model in the
calculation, given by a full line, is compared with two other quantities.
One of them  is:
\begin{equation}
\bar A^i_n \simeq {1 \over d_n} A^{exp,i}_3~,
\label{extr-1}
\end{equation}
where $i$ stands for ``Collins'' or ``Sivers'',
$A^{exp,i}_3$ is the result of the full calculation, 
simulating data, and $d_n$ is the neutron dilution
factor. The latter quantity is defined as follows, for a neutron $n$
(proton $p$) in $^3$He:
\begin{eqnarray}
d_{n(p)}(x,z)=
{\sum_q e_q^2
f^{q,{n(p)}} 
\left ( x \right )
D^{q,h} \left ( z  \right )
\over
\sum_{N=p,n}
\sum_q e_q^2
f^{q,N} 
( x )
D^{q,h} 
\left ( z \right )
}
\label{dilut}
\end{eqnarray}
and, depending on the standard parton
distributions, $ f^{q,N} ( x )$,
and fragmentation functions, $D^{q,h} 
\left ( z \right )$, it
is experimentally known (see \cite{mio} for details). 
$\bar A^i_n $ is given by the dotted curve in the figures.
The third curve, the dashed one, is given by 
\begin{equation}
A^i_n \simeq {1 \over p_n d_n} \left ( A^{exp,i}_3 - 2 p_p d_p
A^{exp,i}_p \right )~,
\label{extr}
\end{equation}
i.e. $^3$He is treated as a nucleus
where the effects of its complicated
spin structure, leading to a depolarization
of the bound neutron, together with the ones of
Fermi motion and binding, can be taken care
of by parameterizing the nucleon effective polarizations,
$p_p$ and $p_n$.
One should realize that
Eq. (\ref{extr-1}) is the relation which should hold
between the $^3$He and the neutron SSAs if there were no nuclear effects,
i.e. the $^3$He nucleus were a system of free nucleons in a pure $S$ wave.
In fact, Eq. (\ref{extr-1}) can  be obtained from Eq. (\ref{extr}) by 
imposing $p_n=1$ and $p_p=0$.
It is clear from the figures that the difference 
between the full and dotted curves,
showing the amount of nuclear effects, is sizable,
being around 10 - 15 \% for any experimentally relevant $x$ and $z$,
while the difference between the dashed
and full curves reduces drastically
to a few percent, showing that the extraction
scheme Eq. (\ref{extr}) takes safely into account
the spin structure of $^3$He, together with Fermi
motion and binding effects. 
This important result is due to the peculiar kinematics
of the JLab experiments, which helps in two ways.
First of all, to favor pions from current fragmentation, 
$z$ has been chosen in the range $0.45 < z < 0.6$,
which means that only high-energy pions are observed.
Secondly, the pions are detected in a narrow cone around the direction
of the momentum transfer. As it is explained in \cite{mio},
this makes nuclear effects in the fragmentation 
functions rather small. The leading nuclear effects are then 
the ones affecting the parton distributions, already found
in inclusive DIS, and can be taken into account
in the usual way, i.e., using Eq. (\ref{extr}) for the extraction of the
neutron information. In the figures,
one should not take the 
shape and size of the asymmetries too seriously,
being the obtained quantities 
strongly dependent on the models chosen for the unknown distributions
\cite{model}.
One should instead consider the difference between
the curves, a model independent
feature which is the most relevant outcome of the present
investigation. 
The main conclusion is that Eq. (\ref{extr}) will be a valuable tool
for the data analysis of
the experiments \cite{prljlab,jlab12}.

While the analysis of Ref. \cite{mio} has been performed
assuming the experimental set-up of the experiment
described in Ref. \cite{prljlab}, but using DIS kinematics, a
further analysis is being carried on 
\cite{new},
to investigate possible
nuclear effects related to the finite values of the momentum
and energy transfers, $Q^2$ and $\nu$, in the actual experiment.
In Ref. \cite{new}, the description of Ref. \cite{mio} 
will be also improved, implementing a relativistic Light
Front treatment to evaluate the nuclear
polarized spectral function.
Besides, the problem of possible effects beyond IA, such as 
final state interactions, will be addressed, 
and more realistic models of the nucleon structure, able to
predict reasonable figures for the experiments, will be
included in the general scheme. 


\section*{Acknowledgments}
I thank the organizers of the Conference
for the kind invitation,
A. Courtoy, F. Fratini
and V. Vento
for a fruitful and pleasent collaboration,
A Del Dotto and G. Salm\`e for
relevant discussions. 
This work is supported in part by the INFN-CICYT agreement.



\section*{References}

\begin{thebibliography}{9}

\bibitem{bdr}
  Barone V, Drago A and Ratcliffe P
 2002 {\it Phys.\ Rept.\ }  {\bf 359} 1 

\bibitem{ferrara}
Ciullo G, Contalbrigo M, Hasch D, Lenisa P 2009
{\it Proc. of the Second Workshop on Transverse
Polarizatin Phenomena in Hard Processes, ``Transversity 2008'',
Ferrara, Italy, 28-31 May 2008}, (Singapore: World Scientific).

\bibitem{sidis}
Bacchetta A, Diehl M, Goeke K, Metz A, Mulders P.J. and Schlegel M 2007
{\it J. High Energy Phys.}
JHEP {\bf 0702} 093

\bibitem{D'Alesio:2007jt}
D'Alesio U and Murgia F 2008
{\it Prog.\ Part.\ Nucl.\ Phys.\ }  {\bf 61} 394
  
\bibitem{Pasquini:2008ax}
 Pasquini B, Cazzaniga S and Boffi S 2008
{\it Phys.\ Rev.\ }  D {\bf 78} 034025

\bibitem{Mulders:1995dh}
Mulders P.J. and Tangerman R.D. 1996
{\it Nucl.\ Phys.\ }  B {\bf 461} 197
  [Erratum-ibid.\  1997 B {\bf 484} 538]
  
\bibitem{Cahn:1978se}
Cahn  R.~N. 1978
{\it Phys.\ Lett.\ }  B {\bf 78} 269

  \bibitem{sivers}
Sivers D.W. 1990
{\it  Phys.\ Rev.\ }  D {\bf 41} 83; 
1991 {\it  Phys.\ Rev.\ }  D {\bf 43} 261 
  
\bibitem{Boer:1997nt}
Boer D. and Mulders P.J. 1998
{\it  Phys.\ Rev.\ }  D {\bf 57} 5780

\bibitem{Brodsky:2002cx}
Brodsky S J, Hwang D S and Schmidt I 2002
{\it Phys.\ Lett.\ }  B {\bf 530} 99
  
  \bibitem{trento}
Bacchetta A, D'Alesio U, Diehl M and Miller C A 2004
{\it Phys.\ Rev.\ }  D {\bf 70} 117504 
  
\bibitem{brodhoy}
Brodsky S J, Hoyer P, Marchal N, Peigne S and Sannino F 2002
{\it  Phys.\ Rev.\ }  D {\bf 65} 114025 

\bibitem{coll2}
Collins J C 2002
{\it  Phys.\ Lett.\ }  B {\bf 536} 43

 \bibitem{jiyu}
Belitsky A V, Ji X and Yuan F 2003
{\it Nucl.\ Phys.\ } B {\bf 656} 165 

\bibitem{yuan}
Yuan F 2003
{\it Phys.\ Lett.\ }  B {\bf 575} 45
  
\bibitem{Gamberg:2007wm}
Gamberg L.P., Goldstein G.R. and Schlegel M 2008
{\it Phys.\ Rev.\ }  D {\bf 77} 094016 

\bibitem{Bacchetta:2008af}
Bacchetta A, Conti F and Radici M 2008
{\it  Phys.\ Rev.\ }  D {\bf 78} 074010

\bibitem{Pasquini:2010af}
Pasquini B and Yuan F 2010
{\it Phys.\ Rev.\ }  D {\bf 81} 114013

\bibitem{baro}
Barone V, Melis S and Prokudin A 2010
{\it  Phys.\ Rev.\ }  D {\bf 81} 114026

\bibitem{boernew}
Boer D 2011
On a possible node in the Sivers and Qiu-Sterman functions {\it Preprint}
arXiv:1105.2543 [hep-ph]

\bibitem{bacnew}
Bacchetta A and Radici M 2011
Constraining quark angular momentum through semi-inclusive measurements
{\it Preprint}
arXiv:1107.5755 [hep-ph]

\bibitem{Courtoy:2008vi}
Courtoy A, Fratini F, Scopetta S and Vento V 2008
{\it  Phys.\ Rev.\ } D {\bf 78} 034002


\bibitem{Courtoy:2008dn}
Courtoy A, Scopetta S and Vento V 2009
{\it  Phys.\ Rev.\ } D {\bf 79} 074001

\bibitem{Courtoy:2009pc}
Courtoy A, Scopetta S and Vento V 2009
   {\it Phys.\ Rev. }  D {\bf 80} 074032

\bibitem{Jaffe:1974nj}
Jaffe R L 1975
  {\it Phys.\ Rev. }  D {\bf 11} 1953

\bibitem{ik}
  Isgur N and Karl G 1978
  {\it Phys.\ Rev. }  D {\bf 18} 4187;
  1979 {\it Phys.\ Rev. }  D {\bf 19} 2653
  [Erratum-ibid.\ 1981  D {\bf 23} 817 ]
  
\bibitem{Traini:1997jz}
  Traini M, Mair A, Zambarda A and Vento V 1997
{\it  Nucl.\ Phys.\ }  A {\bf 614} 472
  
\bibitem{coll3}
  Efremov A V, Goeke K, Menzel S, Metz A and Schweitzer P 2005
  {\it Phys. Lett.}  B {\bf 612} 233;
  Collins J C, Efremov A V, Goeke K, Menzel S, Metz A and Schweitzer P 2006
  {\it Phys.\ Rev. }  D {\bf 73}, 014021 


\bibitem{Courtoy:2011mf}
Courtoy A Scopetta S and Vento V 2011
{\it Eur.\ Phys.\ } J.\  A {\bf 47} 49

\bibitem{Ceccopieri:2007ek}
Ceccopieri F A and Trentadue L 2008
  {\it Phys. Lett.}  B {\bf 660} 43 

\bibitem{Kang:2008ey}
Kang Z B and Qiu J W 2009
 {\it Phys.\ Rev. }  D {\bf 79} 016003

\bibitem{Zhou:2008mz}
Zhou J Yuan F and Liang Z T 2009
  {\it Phys.\ Rev. }  D {\bf 79} 114022

\bibitem{Vogelsang:2009pj}
Vogelsang W and Yuan F 2009
  {\it Phys.\ Rev. }  D {\bf 79} 094010 

\bibitem{Cherednikov:2010uy}
  Cherednikov I O, Karanikas A I and Stefanis N G 2010
 {\it Nucl.\ Phys.\ }  B {\bf 840} 379
  \bibitem{Burkardt:2004ur} 
  Burkardt M 2004
  {\it Phys.\ Rev. }  D {\bf 69} 091501;
  {\it Phys.\ Rev. }  D {\bf 69} 057501
\bibitem{hermes} Airapetian A {\sl et al.} [HERMES Collaboration]
2005
{\it Phys. Rev. Lett.} {\bf 94} 012002
\bibitem{compass}
Alexakhin V Y {\sl et al.} [COMPASS Collaboration]
2005
{\it Phys. Rev. Lett.} {\bf 94}, 202002
\bibitem{bro} Brodsky S J and Gardner S 2006 
{\it Phys. Lett.} B {\bf 643} 22  
\bibitem{prljlab} 
Qian X {\it et al.} 2011
Single Spin Asymmetries in Charged Pion Production
from Semi-Inclusive Deep Inelastic on a Transversely
Polarized $^3$He Target {\it Preprint} arXiv:1106.0363v3 [nucl-ex]
\bibitem{jlab12} Gao H {\it et al.} 2011 
Eur. Phys. J. Plus {\bf 126}, 2 
\bibitem{alessio} Del Dotto A., Master Thesis, Universit\`a di Roma
``La Sapienza'', 2011 (unpublished).
\bibitem{mio} Scopetta S. 2007
{\it Phys. Rev. } D {\bf 75} 054005
\bibitem{pisa} 
Kievsky A, Viviani M and Rosati S 1994
{\it Nucl. Phys. } A {\bf 577} 511 
\bibitem{av18} 
Wiringa R B, Stocks V G J and Schiavilla R 1995
{\it Phys. Rev. } C {\bf 51} 38 
\bibitem{over} 
Kievsky A, Pace E, Salm\`e G, and Viviani M 1997
{\it Phys. Rev. } C {\bf 56} 64  
Pace E, Salm\`e G, Scopetta S and Kievsky A 2001
{\it Phys. Rev. } C {\bf 64} 055203 
\bibitem{ans}
Anselmino M, Boglione M, D'Alesio U, Kotzinian A, 
Murgia F and Prokudin A 2005
,  
{\it Phys.\ Rev. }  D {\bf 71} 074006;
2005 {\it Phys.\ Rev. }  D {\bf 72} 094007
[2005 Erratum-ibid.\  D {\bf 72} 099903].
\bibitem{model} 
Amrath D, Bacchetta A and Metz A 2005 {\it Phys. Rev.} D {\bf 71} 
114018 
\bibitem{old} Ciofi degli Atti C, Scopetta S, Pace E 
and Salm\`e G 1993
{\it Phys. Rev.} C {\bf 48} 968 
\bibitem{new} Del Dotto A, Salm\`e G and Scopetta S, in preparation.

  
\end{thebibliography}
\end{document}